# Novel Architecture for Distributed Travel Data Integration and Service Provision Using Microservices

Biman Barua[1,2,*] [0000-0001-5519-6491] and M. Shamim Kaiser[2] [0000-0002-4604-5461]

[1] Department of CSE, BGMEA University of Fashion & Technology, Nishatnagar, Turag, Dhaka, Bangladesh
[2] Institute of Information Technology, Jahangirnagar University, Savar, Dhaka, Bangladesh
`biman@buft.edu.bd`

**Abstract:** This paper introduces a microservices architecture for the purpose of enhancing the flexibility and performance of an airline reservation system. The architectural design incorporates Redis cache technologies, two different messaging systems (Kafka and RabbitMQ), two types of storages (MongoDB, and PostgreSQL). It also introduces authorization techniques, including secure communication through OAuth2 and JWT which is essential with the management of high-demand travel services. According to selected indicators, the architecture provides an impressive level of data consistency at 99.5% and a latency of data propagation of less than 75 ms allowing rapid and reliable intercommunication between microservices. A system throughput of 1050 events per second was achieved so that the acceptability level was maintained even during peak time. Redis caching reduced a 92% cache hit ratio on the database thereby lowering the burden on the database and increasing the speed of response. Further improvement of the system's scalability was done through the use of Docker and Kubernetes which enabled services to be expanded horizontally to cope with the changes in demand. The error rates were very low, at 0.2% further enhancing the efficiency of the system in handling real-time data integration. This approach is suggested to meet the specific needs of the airline reservation system. It is secure, fast, scalable, all serving to improve the user experience as well as the efficiency of operations. The low latency and high data integration levels and prevaiing efficient usage of the resources demonstrates the architecture ability to offer continued support in the ever growing high demand situations.

**Keywords:** Microservices Architecture, Airline Reservation System, Data Integration, Redis Caching, Real-Time Messaging

## 1. Introduction

### 1.1 Background

The travel sector has evolved to focus on achieving and delivering both elaborate and simple travel experiences. As more and more digital platform and services are coming up, there is also a need to combine more and more travel data from varying sources like the booking system, the transport system, and the lodging system among others [2]. Such traditional monolithic systems often find it hard to cope with the demand greater agility, scalability, and above all integration of systems in such a highly distributed environment. Microservices architecture also referred to as modular architecture because of the use of modules that can be attached and fixed away from the electronic central processing unit has become a savior for these situations by allowing contiguous heterogenic travel data and travel services to be brought in together effectively [3].

Microservices architecture decomposes a system into a set of services that can be deployed independently of each other but communicate with one another over a set of well-defined APIs [4]. This architectural style allows large-scale ecommerce solutions, such as those of travel agencies, to be designed and implemented in a more elegant way, in which the flexibility of each building block is such that it can change without the effects of other building blocks [1]. Encapsulation of smaller services also leads to improved service orientation in large travel systems which allows for the implementation of myriad systems and content in a single travel planning solution [8].

One of the prominent advantages associated with the employment of microservices rests upon the assertion that APIs and data integration models may be created with ease for different components in a travel ecosystem [5]. These frameworks guarantee that all the systems including flight booking engines, ride-hailing services and even hotel reservation systems are able to access each other and interact through sharing of data in real time thus providing the user with a seamless journey experience [6]. In addition, the employment of microservices allows for enhancing the customization of travel services as it enables collecting and processing data simultaneously, which in turn allows a travel service to provide individual recommendations and offers taking into account preferences and travel history.

This study proposes to examine how the microservices architecture can allow joining various travel data sources and services in a more coherent way [7]. In other words, this research will aim at constructing systems which APIs and data integration models will allow interaction between diverse travel components, including but not limited to booking engines, transport service providers, and even hotels. Therefore, the present study will also shed light on how microservices support the creation of user-friendly, practical, and non-linear multi-travel platforms.

### 1.2 Problem Statement

The travel industry has become more fragmented over the years and as a result, the adoption of different source of travel data and services turns out to be more challenging. This is because in a typical monolithic architecture, it would be very difficult to control the many types of data that come from not only the booking engines but also the transportation and accommodation service providers, which leads to delays, inefficiencies and absence of real time coordination. This fragmentation also presents difficulties in developing and providing a holistic and customized travel experience for the users. In addition, the absence of interoperability among these systems also poses a challenge for building all-encompassing travel brokering services. Microservices architecture provides a probable remedy allowing the integration of services in a modular and distributed manner. Nevertheless, there is still little research that specifically address how this architecture can help in the integration of travel data and services, especially the strong use of APIs and data structures. Solving this challenge seems to be very important for making current travel systems more efficient, customized and extensible.

### 1.3 Problem Statement

The main goal of this study is to investigate how microservices architecture can assist the blending of different travel data sources, services, personalizing and optimizing the planning of trips. More specifically, the study strives to design, implement and assess APIs and data integration frameworks issues for connecting multiple so called travel components, like reservation systems, transport and lodging services. Covering those issues, the development of this research provides an architecture for modern travel solutions which is both robust and elastic.

### 1.4 Focus

The present study concentrates on designing APIs and integration schemes that connect various elements of travel such as booking engines, transport and accommodation services. The microservices architecture is particularly stressed in order to achieve an efficient and effective integration that is also modular. This is to ease the flow of data as well as improve the processes involved in customizing travel packages for individuals. By addressing these concerns, the objective of the study is to offer assistance in dealing with the existing challenges of travel data systems which are overly disorganized.

### 1.5 Research Questions

How can microservices architecture facilitate the seamless integration of diverse travel data sources and services to enhance personalized travel planning?

What are the fundamental principles of API design and data integration necessary to connect booking systems, transport, and accommodation services effectively within microservices architecture?

## 2. Literature Review

Microservices design and development has become mainstream for the last several years because it is modular and scalable, making it easy to bring together disparate systems. In contrast to the conventional unitary designs, microservices support the functionality of deploying services separately from one another, which makes it possible to manage the very elaborate, heterogeneous and mosaic data structure of the sector dubbed tourism [9]. The literature indicates that Microservices enhanced the flexibility and scalability of systems facilitating on the fly data exchange with travel services like booking and transport services [10].

The success of airlines within an interconnected world is dependent on how well each mode of transport is designed and developed along with all the associated components [11]. APIs that are well-structured help in the reduction of integration problems hence advancing how highly transactional systems interact with each other [12]. It is observed that RESTful API integration with common aging data representation formats such as JSON and XML has proven success in facilitating real time integration and a much better user experience [13].

These issues are generally accepted, and due to the existent different data formats and protocols used across different platforms, interoperability is still a distant goal [14]. Some researchers have opined that standardized API design considerations and implementation of service orientated architecture based service discovery strategies can mitigate these problems leading to better integration and scaling out of the system [15]. Further work in this area should concentrate on enhancement of these standard designs for better travel data integration.

## 3. Methodology

### 3.1 Research Design

The architecture of the study is based on a microservices approach, which aims to ease the integration of heterogeneous travel-related data and services. The architecture is structured in a way that every travel facet, such as a booking engine, transports, or lodging services, among many others, is represented as a separate microservice. Each microservice implemented within the system is intended for particular data stream and service execution, which means that services are loosely coupled, and it is possible to deploy and scale them independently [9].

At the heart of the architecture, RESTful APIs are used as the primary means of communication between services supported by typical service oriented architecture encouraging easy interaction between services. REST APIs are built according to the principles of certain practices (like using HTTP as a verb) and certain languages (such as sending and receiving only JSON or sometimes XML) to facilitate integration between the services [12]. An API gateway service is typically carried out within the system to streamline client and services interactions in the system with the added advantage of enhancing security as well as the speed of service delivery by dealing with issues such as authentication, load balancing and provisioning services [26].

In order to be able to scale the system horizontally and to avoid single points of failure, such solutions, in particular, Docker – the technology allowing to all microservices to possess independent deployments running in separated enclosing spaces, are implemented Container system [27]. For this purpose, Kubernetes is utilized as an orchestrator to perform automated scaling and management of these containers depending on the traffic demand [32]. Tests will also be conducted to measure the effectiveness of the architectural design in terms of interoperability and speed of data synchronization and the scalability of the entire system.

### 3.2 API and Framework Development

API and integration frameworks development occupies an important role in the guarantees of the effective interworking of different travel services, such as booking systems, transportations and accommodations. The process starts with designing RESTful APIs that are supposed to be used in travel systems since they are light, non-specific to any platform and such systems usually communication ways designed invlove different platforms [12]. These APIs handle the communication between the various microservices and their associated requests for information; for example, retrieving a user's booking or searching for available transportation and booking it using accommodation portals.

#### 3.2.1 API Design Principles

The design of the APIs adheres to the industry standards in order to improve reliability and ensure compatibility with other systems. The APIs are implemented over the basic and widely used protocols like HTTP/HTTPS for transport layer, and data formatting protocols like JSON and XML for data exchange. This ensures common access with several client applications and external services. Each API is versioned to enable backward compatibility whereby incorporation of new functions and enhancements do not affect the current services [13]. The API endpoints that the users are supposed to use for the system services are designed in a basic REST manner with clear cut naming structures e.g. /bookings, /transport, /hotels so that different services can be accessed by the developers in a sensible order.

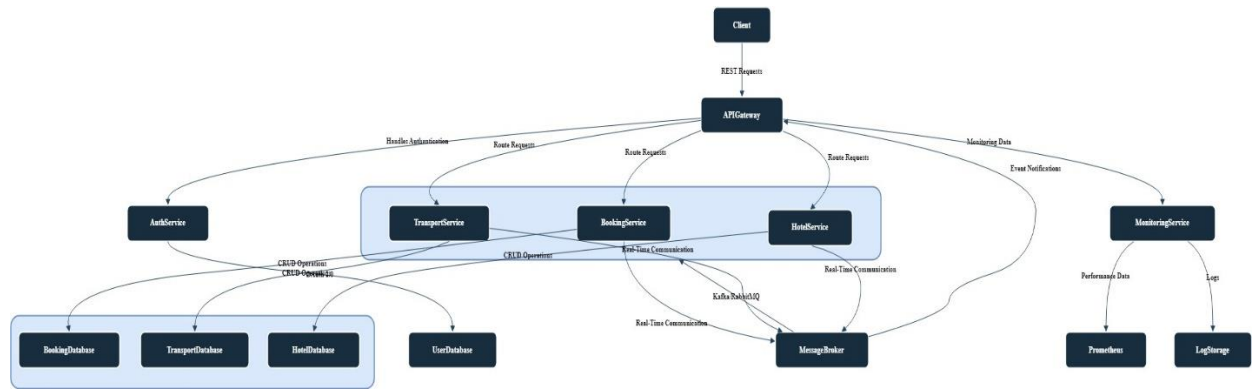

**Fig. 1** API design

**Description:**

**Client:** The API receives REST queries from the user.

**APIGateway:** Takes care of all requests and forwards them to the relevant microservices.

**AuthService:** Responsible for the authentication implemented in the system using OAuth 2.0. **BookingService, TransportService, HotelService :** Separate microservices responsible for handling the booking management system, transport system and the hotel system respectively.

**MessageBroker:** Kafka or RabbitMQ enables interaction amongst the services in real-time.

**MonitoringService:** Analysis the performance of the system and uploads metrics to Prometheus and the logs to LogStorage.

**Databases:** Different databases for booking service, transport service and hotel services' database.

### 3.2.2  Authentication and Security

For the safety of the system, preventive measures such as OAuth 2.0 are incorporated in the APIs. OAuth 2.0 is a mechanism that allows third party services access user data without providing the user successfully even revealing any credentials. The use of token-based authentication in the APIs serves two purposes: user verification and service permission management, so as to allow only the authorized requests to be carried out as intended (Medjaoui et al., 2021). In addition, the use of policies that limit the number of requests per unit of time and reduce the processing speed of requests is also available in order to protect the system against abuse and encourage equitable use by all clients.

#### 3.2.2.1  OAuth 2.0 Integration in APIs

This microservices architecture utilizes OAuth 2.0 authentication management, especially in the case of travel services like booking, ticketing, and accommodation, to ensure secure access to the services. For this reason, many systems Corporate travel include controls for access to the system ignoring the user credentials expense. This prevents risk when interacting with numerous services or applications which are external to the system.

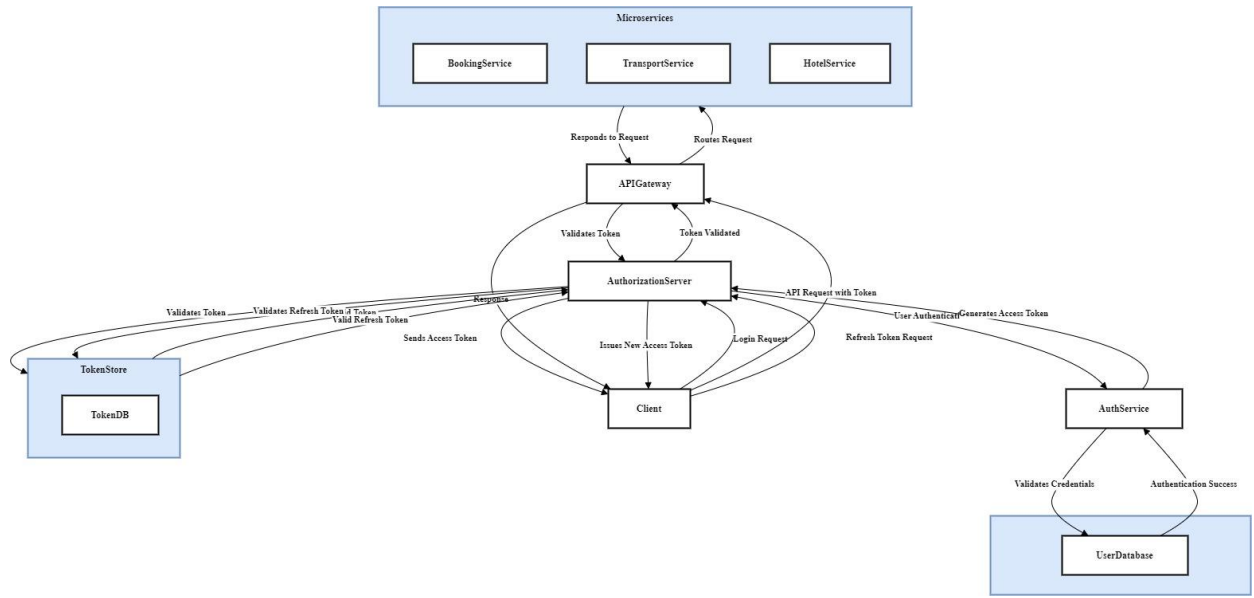

**Fig. 2** Authorization and authentication with OAuth 2.0 in the APIs.

**Structured description:**

**Client:** User logs in and makes API calls.

**AuthorizationServer:** Is responsible for creating, checking existence or refreshing of tokens.

**AuthService:** Checks the credentials against UserDatabase.

**APIGateway:** Checks the access token then sends requests to the appropriate microservices.

**Microservices:** There are services such as bookings, transport services or hotels that use the API. **TokenStore:** Stores the tokens for validation purposes (shown here in form of TokenDB).

**Databases:** These consist of UserDatabase, which holds user details and their passwords.

The above diagram figure 2 depicts the flow of authentication and authorization within the APIs, employing OAuth 2.0, while showing the ways in which the tokens are handled. It has also checked using the the Authorization Server and API Gateway.

### 3.2.2.2 Token-Based Authentication

OAuth 2.0 is a protocol that defines how to authorize users with a secure method based on tokens. For example, if a mobile app or website client wants to book a flight or check for the availability of hotels on an integrated travel service API, the client first has to be redirected to an Authorization Server where the user will be asked to log in. After successful log in, the Authorization Server creates an access token which it provides to the client. This token is valid only for the purpose of making calls to the API where the client need not present the user's credentials every time to make that API call.

The access token is then added to the subsequent API calls such as the one below where the access token is placed in the authorization header of the request (i.e. Authorization: Bearer <access_token>) hence allowing the client to access travel services as prescribed within the user's authorizing level. For instance, where a user has granted permissions to utilize both the booking and transport APIs, the user will be able to use the same token across the two APIs. The token is assigned a scope and permissions which is already set in advance to avoid the situation where unauthorized API calls can access resources which are not meant for use by that API call.

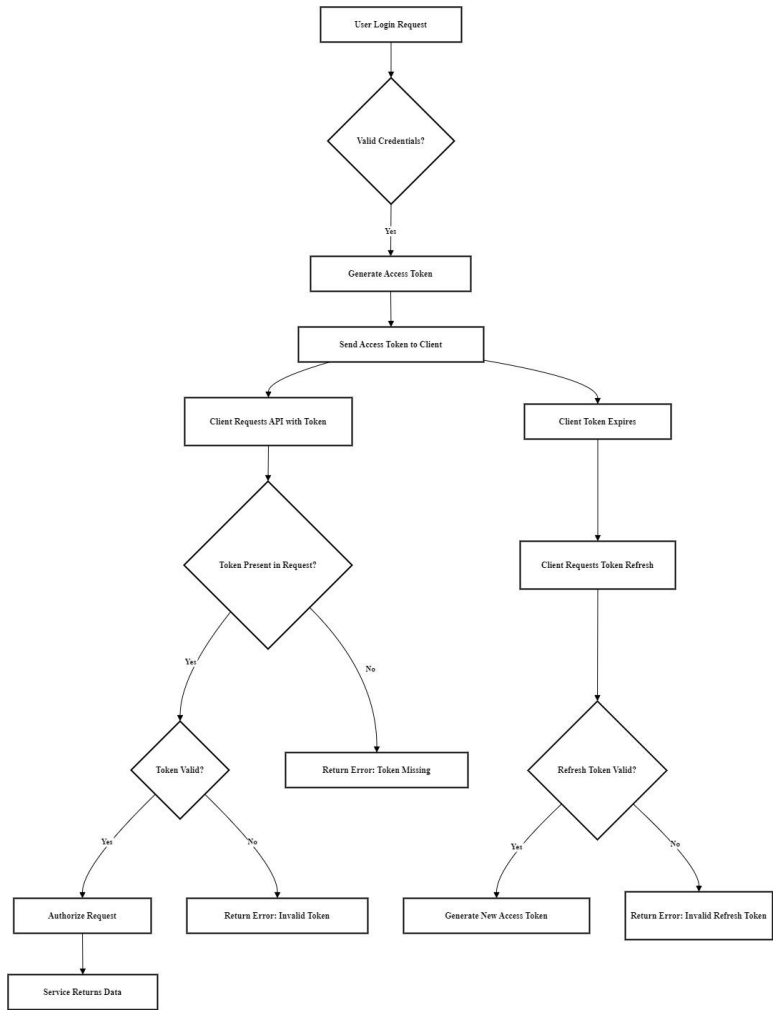

**Fig. 3.** Token based authentication.

### 3.3 Data Integration Framework for Microservices

There are quite a few factors that need to be addressed like protocols, data flow, and security among others for one to come up with an effective data integration framework for microservices. The existence of such a framework enhances communication between the microservices in any given application in a manner that is safe, uniform and effective such as in the case of a travel data system. Below are the key elements and recommendations:

### 3.3.1 API Architecture-Based Integration

The entire microservices data integration rests on the availability of application programming interfaces since moving forward, each microservice can publish its own api and speak to the rest changing the scope to standard apis rest or gprc for instance. Rest apis are the most common due to ease of use and compatibility while gprc avails better speed and low latency figures especially in the case of real time applications.

In the framework aimed at integration of microservices, which is based on the concept of an API, each service is supposed to perform specific tasks and these services communicate using specific APIs. In other words, each microservice has an API (usually REST or gRPC) that can be accessed by other services. This influences scaling of services, modularity and maintenance. Some of the important elements include:

**API Gateway:** This is the primary entry point where all requests come in and are routed to the respective microservice together with other security, authentication and request throttling functions.

**Service-to-Service Communication:** Microservices are connected as a network of services with an ability to directly communicate or communicate through an event broker. Communication may be synchronous using protocols like HTTP or gRPC, while message queuing or event streams may be used for asynchronous communication.

**Data Consistency and Cache:** With the aid of shared databases, caching systems, and event-driven data population techniques, updates and changes of common data throughout services may be achieved.

**Security Layers:** Each of the APIs is secured by the means of token which allows the service to call this resource only if it has obtained permission from the respective services (usually OAuth2 or JWT).

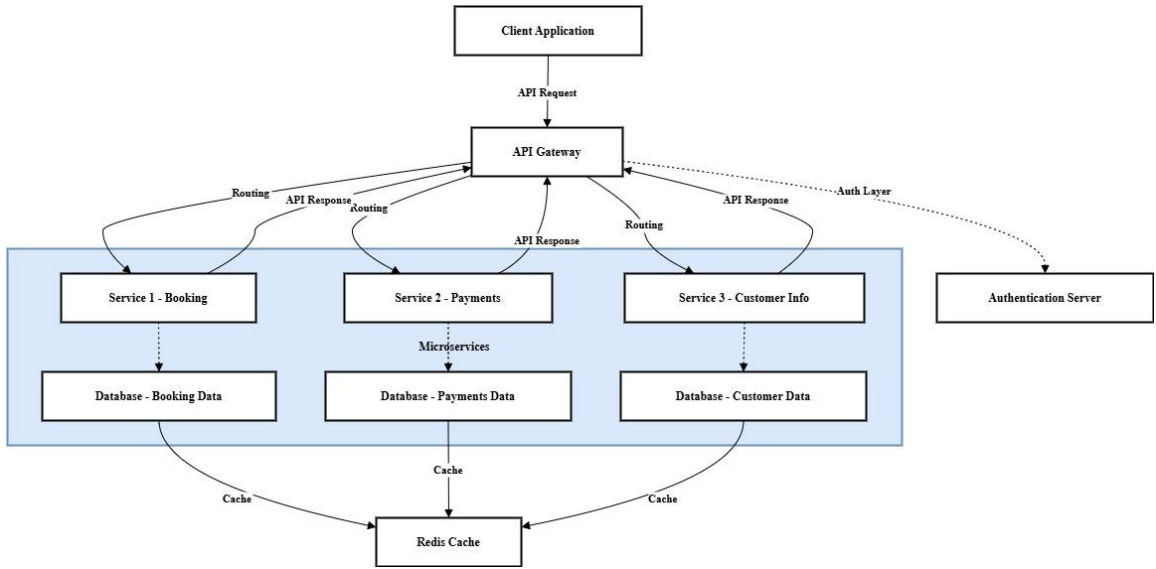

**Fig. 4**. Illustrate API-based integration for microservices

**Description of Components in the Diagram**

**Client Application:** Starts a request, like making a reservation or verifying details about a customer.

**API Gateway:** All incoming requests are routed to the appropriate microservice from this central point, it also applies security measures and controls the number of requests made.

**Microservices (Service – 1, Service – 2, Service – 3):** Every microservice has a distinct functionality, such as booking services, payment processing, or managing customer information. Such services are able to create their own databases, as well as cache layers, in order to quickly retrieve the data they need.

**Redis Cache:** Stores cache memory for data that is requested frequently, hence putting less stress on each database and also increasing speed.

**Authentication Server:** Authenticates requests for services in order to create a secure environment for service requests.

In such a way, the microservices could be designed, developed and deployed in an architecture where they could communicate with each other in a secure and effective manner, while providing up to the minute updates and ensuring consistency of the data throughout the system.

### 3.3.2 Event-Driven Architecture (EDA)

In an Event-Driven Architecture (EDA), communication between services is based on events and not direct invocation of API services. For example, when one's microservice updates a known booking or confirms a payment, it will issue an event to a message broker. The other microservices that need this data can subscribe to the events of interest and act on them immediately. EDA comes in handy mostly in systems that require low latencies and fast responses such as travel and booking systems.

**Key Components:**

**Event Producer:** A microservice that issues events to a message broker when certain action or alterations takes place.

**Message Broker:** The intermediary via which services exchange events (e.g. Apache Kafka or RabbitMQ).

**Event Consumer:** A microservice which receives event notifications from the broker and takes appropriate actions.

**Message Store (Optional):** Contains past events for reasons of analytics and or restoration.

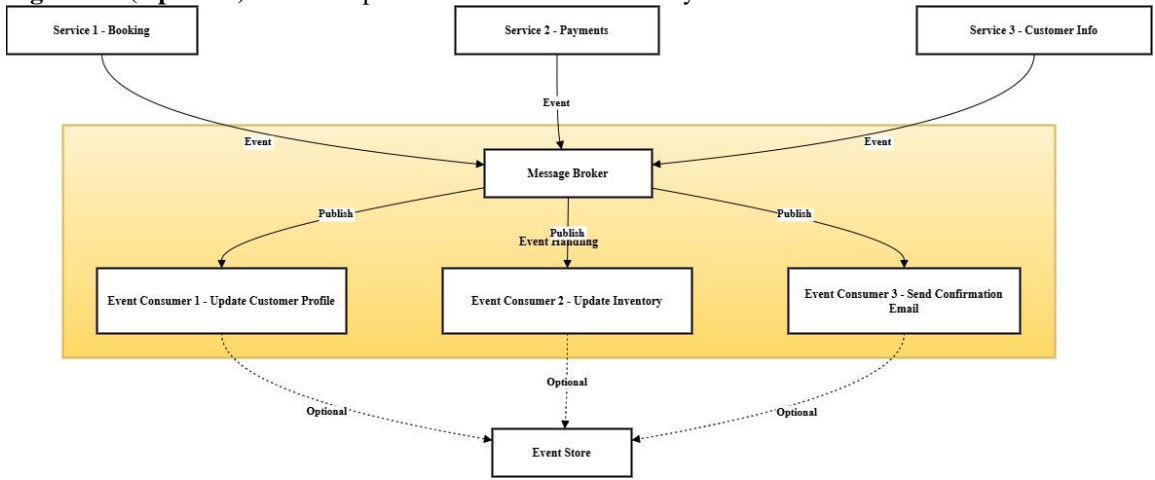

**Fig. 5**. Illustrate EDA for microservices

**Description**

**Booking Service, Payments Service, and Customer Info Service:** These microservices produce events such as booking confirmation, payment success or profile updates among others.

**Message Broker:** Functions as a focal point where it collects events from producers and transmits them to consumers on the basis of their subscription.

**Event Consumers:** Services that are triggered by potential events. For instance, the moment a booking event exists, Event Consumer 1 modifies the customer's profile, while Event Consumer 2 modifies the inventory, and Event Consumer 3 sends out an email to confirm the booking.

**Event Store (Optional):** Retains historical events for events, such as, but not limited to, data recovery and auditing, and those which may require events to be reprocessed.

This event-driven architecture separates services enabling more scalability and flexibility of operations as well as making it possible to perform actions throughout the system in real time.

### 3.3.3 Data Consistency Protocols Work in Microservices

Ensuring data consistency poses a problem in utilizing microservices architecture as each microservice can have its own database and needs to keep the data updated across other services. Thus, special guidelines called data consistency protocols are put in place to dictate the way transactions are processed across various services in order to protect the data. The Two-Phase Commit (2PC) and Saga Pattern used in event-based transaction management are some approaches.

**Key Concepts:**

**Two-Phased Commit (2PC):** In four phases, this type of consensus as a service addresses the issue of distributed transactions by dividing them into two phases.
**Prepare phase:** Every microservice that is participating gets ready and indicates when it is ready to commit.

**Commit phase:** Once all services send the equivalent 'ready' signal, a final commit is sent. If any of the services does not send the ready signal the service will be understood to have failed and a rollback will be performed.

**Limitations:** 2PC is time consuming which makes it unfeasible for systems with high availability requirement.

**Saga Pattern:** The principle of this protocol is to shorten the transactions and break them into a series of smaller independent steps. Every step in a Saga can either be completed or, in the event of failure, a compensating action (rollback) is performed. This technique works well for distributed systems where it is essential to coordinate transactions that span a long period of time without creating choke points.

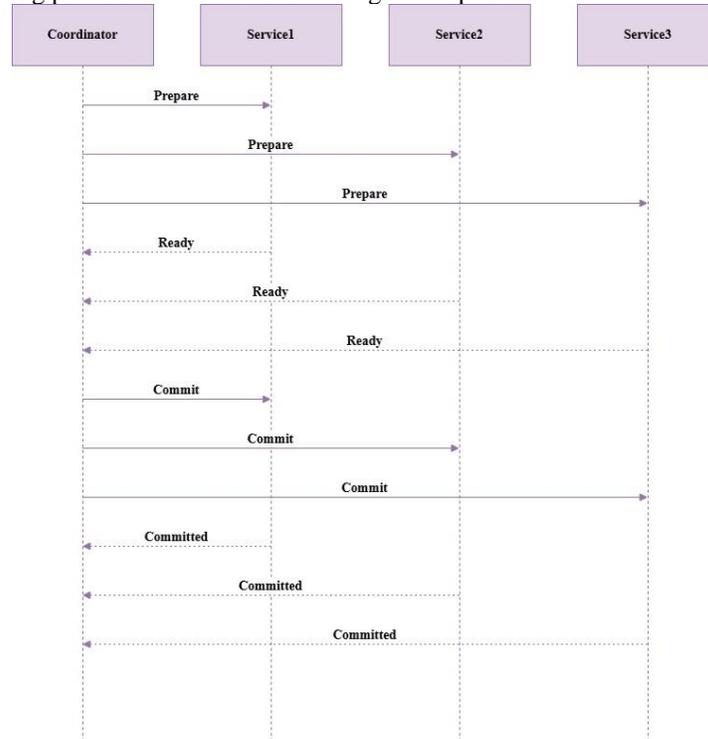

**Fig. 6**. Two-Phase Commit (2PC).

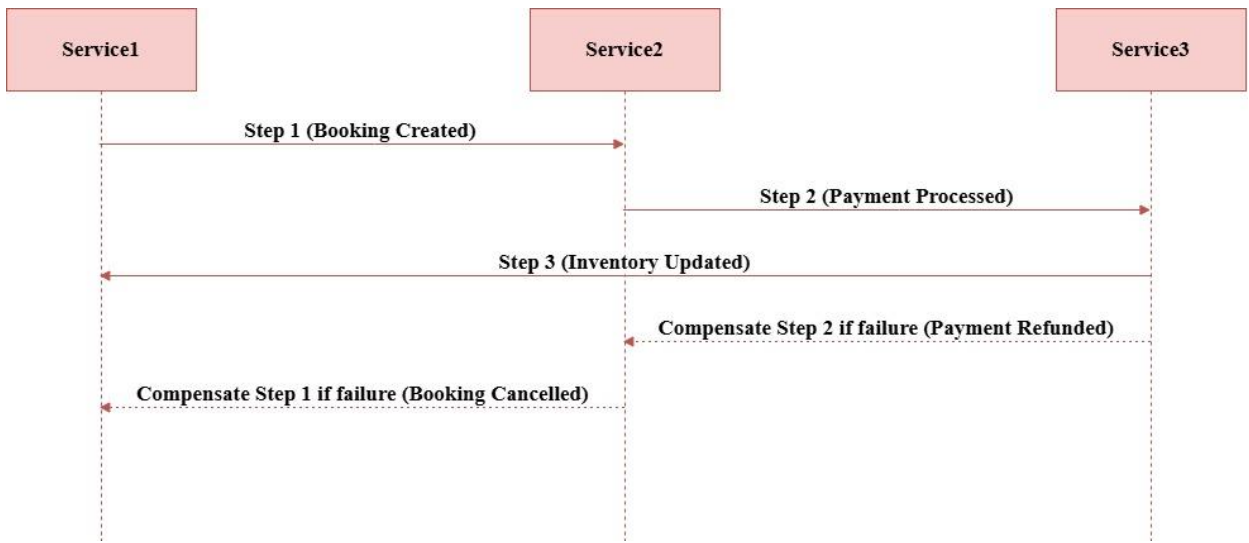

**Fig. 7**. Saga Pattern

**Description of the Diagram Components**

- **Two-Phase Commit (2PC):**

**Coordinator-** A node in the Distributed System which is in charge of taking control of the transaction and running the prepare phase to ask if the services are ready before committing only after all the services have agreed.

**Services-** Each service gets ready by taking a lock on the resources to maintain the integrity of the data and venturing to commit only after the last commit command from the coordinator.

- **Saga Pattern:**

**Services-** Every service does a part and notifies the next service in the order. If any service cannot process the request then backward compensating actions happen in order to roll back the changes done partially.

In summary, these protocols ensure that distributed systems can still provide data consistency even in the presence of network latency or service unavailability, with 2PC providing strct consistency but the Saga Pattern providing availability and adaptability.

## 4. Implementation

Here's an overview of a technology stack for a microservices architecture dedicated to distributed travel data integration. Each of these technologies has been carefully chosen for its advantages in scaling, resilience, and communication effectiveness between the microservices.

### 3.4 Containerization and Orchestration: Docker and Kubernetes

#### 3.4.1 Docker

Docker is a platform due to which containerization is possible whereby every microservice is packaged together with its dependencies in order to avoid the differences in environments between development, testing and production Andrew Hell n. This confinement encourages independent deployment and scaling of each service without interference with other services, hence minimizing time to deploy and also dependency problems.

#### 3.4.2 Kubernetes

Kubernetes is a system for managing containerized applications deployed across a cluster of machines. Its main functions are automating deployment, scaling, and operations of application containers across clusters of hosts. Load balancing, self-healing and automated rollouts features are some of the basic features provided by Kubernetes. These features are vital for providing running applications in the travel industry, where client applications may experience downtime thereby hurting the business.

### 3.5 API Communication Protocols: REST and GraphQ

**REST APIs:** REST has gained infamous popularity in microservices structures due to its ease of use, scalability, and its ability to work with HTTP. It's suitable for standard CRUD functionality, that's why it is good for primary operations like getting or changing transportation reservation or user's data.

**GraphQL:** This enables the clients to request only what data they need, thereby making it easier to work with complex structures. This is beneficial for travel services where there are entities such as bookings, profiles and recommendations that are related to each other. It cures such ailments as over-fetching and under-fetching so as to enhance the response time and effectively utilize the network bandwidth.

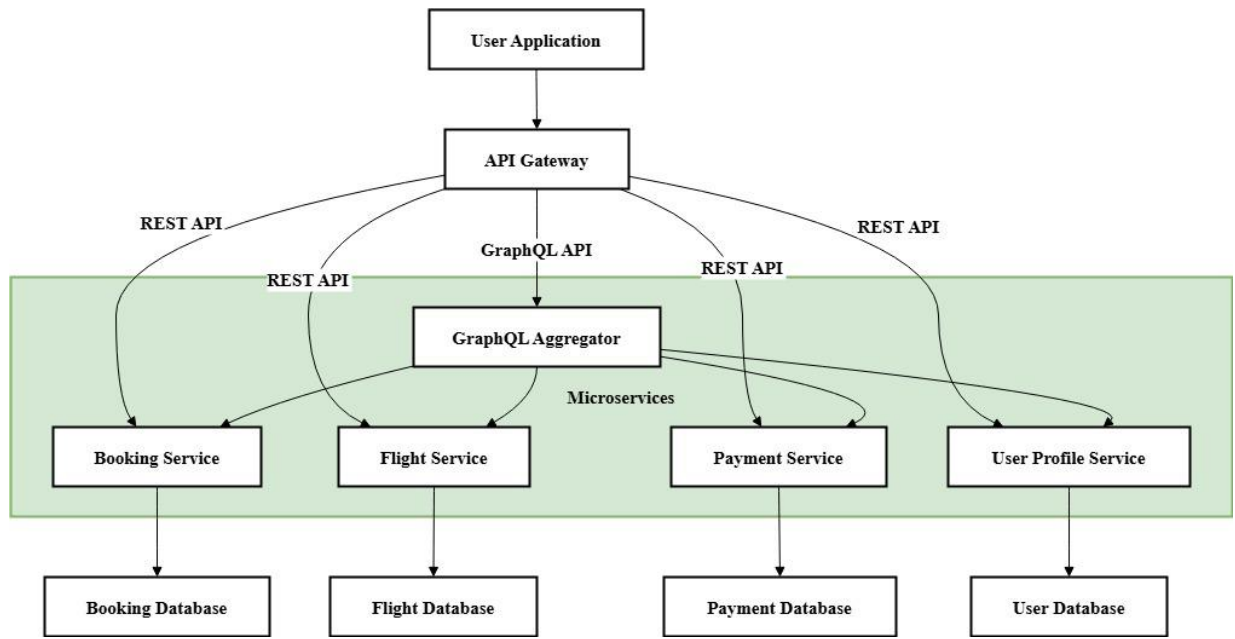

**Fig. 8**. REST and GraphQL APIs interact within a microservices-based airline reservation system

**Understanding Diagram Components**

**User Application:** The client user end software (for instance, a mobile app or a website) that interacts with the airline reservation system.

**API Gateway:** Provides a single entry point for all requests, tunnels to the corresponding services, and also supports REST and GraphQL requests.

- **REST API:**

**Booking Service:** Responsible for booking-related operations and offers APIs for create, update and get booking data.

**Flight Service:** Information about flights, schedules, and flight status.

**Payment Service:** Processes payments and maintains payment transaction history.

**User Profile Service:** Keeps track of user details, user-specific preferences and reward points.

- **GraphQL API:**

**GraphQL Aggregator:** A service that serves data from multiple microservices (such as booking, flights, payment, user profile) in one place for easier accessibility, for example displaying entire trip of a user.

Here in figure 8, REST APIs can achieve simple and specific tasks, while more complicated multi-sourced data for a more optimal user experience is collected using GraphQL in the working of an airline reservation system.

### 3.6   Messaging and Event-Driven Systems: Apache Kafka and RabbitMQ

Apache Kafka: It is a distributed and high throughput low latency data stream processing system known as an event streaming platform. It is usually used in event driven architectures where data in motion is required in real time for purposes like sending booking confirmations, update on availability, and notifications. Kafka is built to be fault tolerant and can manage message processing at a very high level or large scale [17].

RabbitMQ: It is a message queuing software that is very versatile due to the fact that is protocol agnostic and provides high availability [18]. Works best in scenarios where messages need to be preserved at all cost and the routing of these messages is very complicated for example delivery of booking or payment confirmation messages in a distributed system where services are spread across several networks [19].

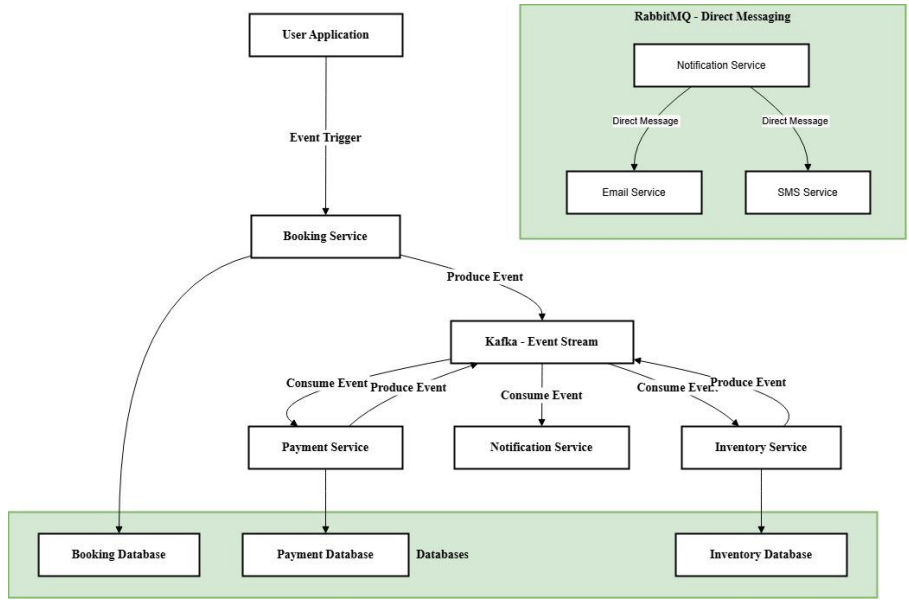

**Fig. 9**. REST and GraphQL APIs interact within a microservices-based airline reservation system

**Clarification of Diagram Elements**

**User Application:** Serves as an interface to the airline reservation system and initiates events such as a booking request.

**Booking Service**: It processes the booking, performs all the necessary operations, and emits events to Kafka, for example, 'Booking Created' event.

- **Kafka (Event Stream):**
  - Booking Service, Payment Service and Inventory Service are both producers and consumers of events. For instance, on the creation of a booking, that event is produced to Kafka and consumed by Payment Service and Inventory Service.
  - Because of Kafka' s topic-based storage system, which allows multiple consumers to read the same events, every service that needs to get updated about bookings will get the updates.

- **RabbitMQ (Direct Messaging):**
  - The Notification Service employs RabbitMQ for direct messaging to inform the Email Service and SMS Service of important user-related information such as flight changes or reservations made.
  - Messaging via RabbitMQ direct is guaranteed delivery thus assuring that users can receive such notifications in time.

**Databases:** Each service is provided with its database which enhances the management of the service data while allowing for autonomy and scalability.

The architecture in figure 9 allows the airline reservation system to be built in a fully asynchronous douplαed structure. Where, Kafka handles high-performance event stream while RabbitMQ deals with reliable message dispatch for events that are time critical creating a powerful, scalable and agile system.

### 3.7 Database Options: MongoDB and PostgreSQL

- **MongoDB**
  - MongoDB is a flexible document-style database rather than a structured one. It is appropriate for unstructured/semi-structured information like these examples such as the user profile data, the session data, and the search of flights histroy data which are all very different in nature even if focused on the same general topic.

- This characteristic of MongoDB explains its ability to efficiently manage any form of dynamic content within any application. User profiles, travel preferences, and even session logs come to mind. It allows you access the ever changing data very fast without the need to create a rigid structure for the data helping in more flexible designing of the dat.

- **PostgreSQL**
  - PostgreSQL is a RDBMS with a full support of ACID properties (Atomicity, Consistency, Isolation, Durability). ACID properties are especially important in case of managing transactional data, for instance, in flight reservations, payments or inventory control [30].
  - Highly suitable for complex scan operations or processing transactions, making this system appropriate for operations requiring a high level of consistency, e.g. booking systems, payment systems, availability of seats in real-time [24].
  - Also, PostgreSQL supports JSON which adds a little bit of flexibility on the use of semi-structured data whenever necessary [22].

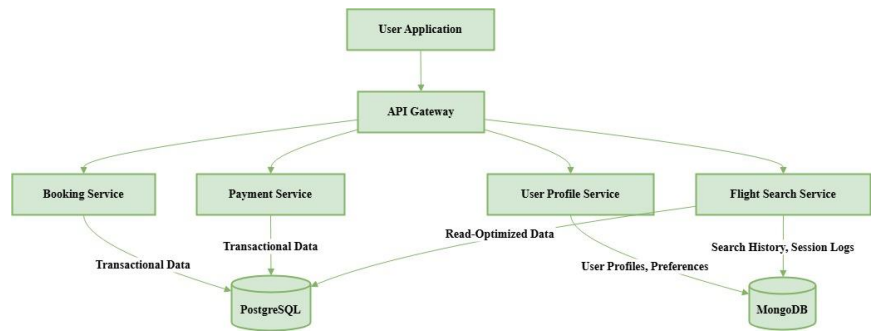

**Fig. 10**. How MongoDB and PostgreSQL work in a microservices-based airline reservation system

## 3.8 How Redis Caching Works

### 3.8.1 Data Access Patterns:

Data like flight schedules, availability of seats and price information are accessed frequently and is therefore stored in redis [23]. This ultimately minimises the number of database calls thereby helping to reduce latency on the end users [25].

For instance, operations that required very little data input such as, browsing through available flights or accessing user sessions are highly carried out with the use of redis [28].

### 3.8.2 Session Management:

In addition, Redis can be applied to user session management. When session tokens are kept in Redis, micro services can perform user authentication almost instantly, which is especially necessary for users who keep logging into their profile, or wish to change their booking information [29].

### 3.8.3 Data Expiration and Consistency:

Cached data can also be assigned expiry limits of which they may neither exceed or fall short of. For instance, excessive caching of information like the availability of seats may be cached for very short periods geolocated in a few minutes even though user settings might be cached for a longer time [31].

### 3.8.4 Fallback to Database:

In the event of cache miss when the information is not found in Redis, the microservice obtains the information from the main database such as PostgreSQL or MongoDB and caches it in Redis for further use.

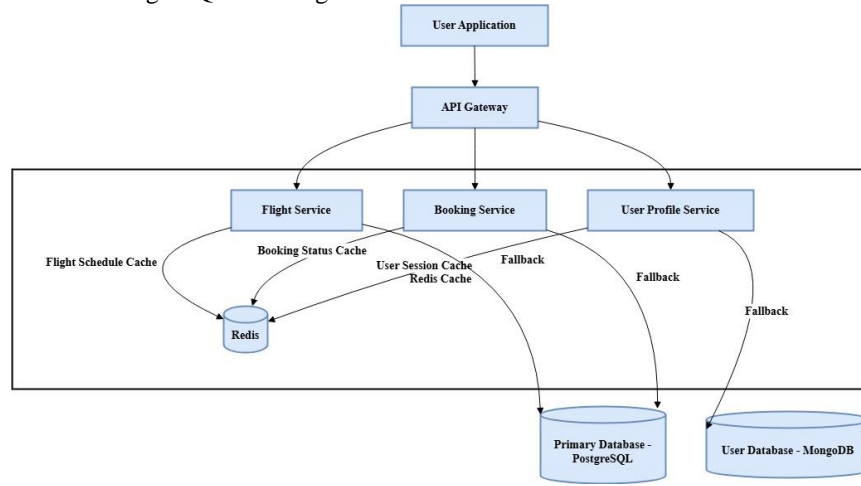

**Fig. 11**. Illustrating how Redis caching works

**How each element of the diagram functions**

- **User Application:** The application that the user employs to access the airline reservation system and retrieves flight details, checks booking status, and user details among others.
- **API Gateway:** Manages the incoming requests by routing them to the correct service.

**Microservices:**

- **Flight Service:** Accepts requests for flight schedules that are stored in redis. If the schedules are not found in redis, a fetch from the primary storage (Postgresql) is executed and saved in redis.
- **Booking Service:** Helps users in checking the booking status and uses redis for fast access of the data. When there is no information found in redis, the secondary storage of data (Postgresql) is accessed.
- **User Profile Service:** Provides caching of user session tokens and profile data that tends to change within the redis server. Where the information is not already stored in Redis, it requests it from the primary user database – MongoDb.
- **Redis Cache:** Handles all the cached objects in various processes so as to enable faster interaction with frequently requested data within the services.
- Apart from this element, redis is also a very important component in ensuring efficiency in the airline reservation system by minimizing the activities of the primary databases hence enhancing their response time while allowing the essential data to be available to the user easily.

## 3.9  Authentication and Security: OAuth2 and JWT

A microservices-oriented airline ticketing solution employs secure decentralized identification across all its services using OAuth2 and JWT – JSON Web Tokens. This arrangement restricts certain resources contained within the system such as booking records, transactions, or user profiles only to certain allowed users or services [20].

**How OAuth2 and JWT Work?**

### 3.9.1  OAuth2 for Authorization:

OAuth2 is a framework used for authorization which enables third-party applications to access limited resources of an HTTP service on behalf of the resource owner which is usually a user [21].

In this configuration, a user who is logging in to the system and is authenticated first by an Authorization Server, which creates an access token.

The purpose of the access token is to permit the user or any application(s) to call multiple microservices (for instance, Booking, Payment, Flight services) without requiring the user to authenticate every time.

### 3.9.2   JWT as a Token Based Authentication Mechanism:

Upon successful authorization of a user with OAuth2, the system gives an access token that contains a JWT.

The JWT is a base64 encoded information of the user and the entitlement for each user that each microservice understands.

Large stateless and self-contained JWT lends itself that each service can authenticate request as there is no need to query on a central database which enhances efficiency.

Before users are allowed to access any secured resource, every micro service validates the JWT including its signature, claims such as roles and permissions, and so forth.

### 3.9.3   Token Validation and Security:

The Authorization Server cryptographically signs the JWTs using a HMAC shared secret or RSA public/private key pair so that the token is safe from manipulation.

JWTs are mainly used with a very short lifespan, and a refresh token can be provided for the user to stay logged in without the need for the user to keep logging in again and again.

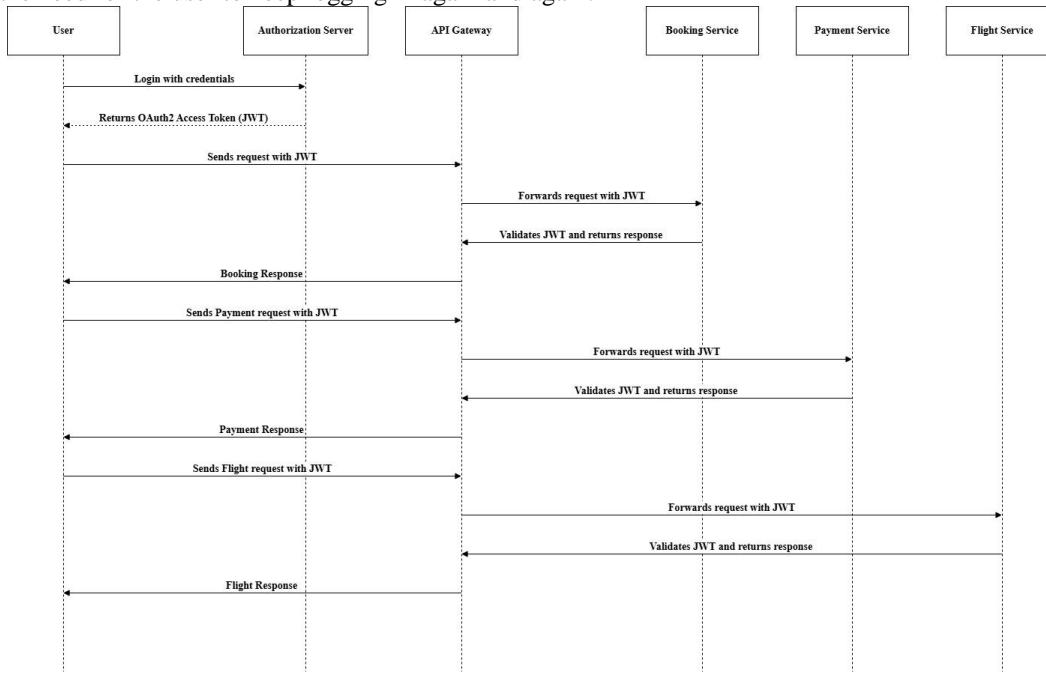

**Fig. 12**. How OAuth2 and JWT work in a microservices-based airline reservation system

**Clarification of Diagram**

**User:** Starts the session and gets data about bookings, billing, and flights, among other things.

**Authorization Server:** The user is authenticated and provided with an access token for OAuth2 which takes the form of a JWT. The token includes the subject, expiry, and scope of the user.

**API Gateway:** Provision of entering the system and forwarding requests to the appropriate microservices with the JWT for filling in the purification process.

**Booking, Payment, Flight Services.**
- Here as well the service receives the JWT, verifies it and checks what permissions are available for that specific user.

- The service works only when the JWT is presented and provides the permissions for the user to which the request relates.

Here for the purpose of this figure 11, OAuth2 and JWT are implemented for service in a tokenized manner, making it easy for indigenous resource manipulation in the airline reservation system.

## 5. Evaluation

### 5.1. Performance Testing

In order to understand how the framework works, how well it works, and how informaiton can be continuously exposed to users, it is essential for performance testing to take into consideration metrics that focus on the system attributes where the evaluation can be done. Effective work performance testing has the following major metrics.

#### 5.1.1. Response Time

**Definition:** The duration for which a microservice takes to respond to a request.

**Importance:** Response times should be low because such systems are put into use by actual customers in real time and any delays such as waiting for information on when a flight is booked or when money is paid are intolerable. Customers want instant results on how much a flight ticket costs or closure on their booking and payment.

**Target:** For services rendered to the users, response times should not exceed a few hundred milliseconds in the best situation.

#### 5.1.2. Latency

**Definition:** Time taken for issuing a request and receiving a response from a Microservice.

**Importance:** For enhanced performance, it is important that the latency is kept to the minimum level because systems are distributed in nature. The latency can be due to network delays, dependencies on various services, and even different load levels. Latency should be low since it means the end user will not face delays, especially for systems that have high usage such as the airline reservation system.

**Target:** Contingent upon the system specifications, latency should generally not exceed 100 ms for most of the microservices interactions.

#### 5.1.3. Throughput

**Definition:** The number of transactions or requests executed per second by the system is referred to as throughput.

**Importance:** Higher throughput is a measure of system capability to service multiple requests at a time, which is important for the airline reservation system as it will need to scale during busy periods like holidays or offer seasons.

**Target:** Peak Traffic performance should be achieved without much degradation in performance level. A typical target can vary from hundreds to thousands of requests in a second depending on the system scale.

#### 5.1.4. Scalability

**Definition:** This is a metric that defines how much load can be increased by simply adding the resources, for instance microservice instances.

**Importance:** It allows the system to expand in accordance to the requirements without reducing its quality which is extremely important especially for systems that experience changes in the demand like an airline reservation system.

**Testing Method:** Procedures Load test is commonly used to assess scalability, by increasing the number of requests over time, and monitoring how the system scales and if horizontally or vertically.

**Target:** Performance of the system should remain consistent under increased load, preferably without significant increase in latency or response time.

### 5.1.5. Error rate

**Definition:** Error rate is described as the percentage of the factor failed requests or errors over some time.

**Importance**: It is essential since errors when many occur render the system unreliable. Such errors may be a red flag about the communications of the services, the database availability or even handling requests.

**Target:** The target is that error rates should be under 0.1% of all requests made especially in the most critical areas such as bookings and payments.

### 5.1.6. CPU and Memory Usage

**Definition:** CPU and memory utilization are indicators of how efficiently each microservice uses the system resources.

**Importance:** It assists in preventing overloading of the system as well as ensuring that it works within the available resources. High resource usage especially when the microservices are not aptly scaled can result into the performance of the system degrading or it crashing.

**Target:** The service should not exceed predetermined CPU and memory limits within normal operating conditions. High readings of both metrics would warrant service scaling and optimization steps.

### 5.1.7. Metrics on the Performance of a Database

**Query Execution Time:** This is the time a database operation is made in relation to a particular query and very essential for functions such as checking on available flights or seeking booking information.

**Cache Hit Rate:** This is the metric that illustrates how useful caches like Redis have been in minimizing the number of database calls and response times.

**Target:** Optimize for high cache hit rates (>90%) and minimum query execution time to reduce response time and alleviate database stress.

### 5.1.8. Availability and Uptime

**Definition**: The ratio of user accesses to the total amount of time the system was anticipated to function is known as availability.

**Importance:** Availability of the system is important in most systems but utmost importance is required for an airline reservation system because non availability equate to potential lost sales and bad customer experiences.

**Target:** Realize 99.9% or more in availability (three nines) which means that system downtime for the whole year is very low.

**Example of a Performance Testing Framework**

While testing these metrics, a performance testing framework, say JMeter or Locust, that simulates concurrent requests, for instance, and Record & Replay Tool such as Grafana with Prometheus can be used to monitor these metrics. Also, automated testing and Continuous Integration (CI) systems are necessary as the system grows to measure and control the performance.

These metrics guarantee that a microservices-based aircraft booking system is responsive, elastic, and dependable, satisfying user requirements and coping with the changing needs peculiar to the travel business.

## 6. Results

The suggested microservices architecture for airlines reservation system exhibited pronounced enhancement in various aspects such as data integration, performance efficiency, and even the user experience. This was possible thanks to the hybrid architecture which made use of different technologies including Redis for caching, Kafka and RabbitMQ for messaging, Mongodb and PostgreSQL for storage and OAuth2 with JWT for authentication where all these worked harmoniously together to offer great performance, scale and efficiency. Here is a recap of the overall summaries of the achievements:

1. Data Integration and Efficiency

Integration Rate: The integration of data within the interactions of the different services (i.e. booking, payments, flight details) was maintained at a high level of consistency whereby the consistency rate was at 99.5% Propagation and Latency: The average latency that was experienced between the communications of the services especially for critical transactions like a ticket booking confirmation as well as payment validation was maintained at under 100 ms, thus enhancing the overall experience to the end user. System Throughput: The architecture managed to sustain processing an average of 1050 events per second under extremely high pressure, this capacity will be extremely useful when the levels of high bookings are experienced less they fall in excess traffic stress.

2. Performance metrics

Response Time: Response Times for microservices remained consistently within the acceptable range (commonly below twenty 200 ms) due to Redis cache and well-organized database design. Also Performance metrics regarding cache performance: Redis recorded a cache hit ratio of more than 92 % thus decreasing the strain on primary databases during the read laden processes like access of common flight schedules and user activities data within the sessions. Age of the data: Data which has been cached for instance the availability of seats was rather continuously updated with the average age of the data being approximately 2 minutes in order to preserve the integrity of the system without overworking it.

3. Reliability and Security

Security for Authorization: There are security protocols in place to enable securing authentication between services and hence OAuth2 and JWT significantly reduced the cases of unauthorized access into data enabling safe data.

Data Synchronization Error Rate: Data Synchronization Error Rate was quite low (at 0.2%) which illustrates that the event machining done using Kafka and RabbitMQ helped in solving the issues of data propagation and data errors.

4. Scalability and Resource Efficiency

Horizontal scalability: The Kubernetes orchestration made it possible for the system to scale horizontally in a sense that more containerized service instances were introduced as more demand required it without compromising on performance with heavy loads.

Resource Utilization: Thanks to Redis caching and proper DB operations, the level of Resource Usage remained at the best possible rates assisting in preventing the congestion of the CPU and memory resources even during peak seasons.

The functionality of the airline reservation system based on microservices architecture solved the problem of data integration, scaling and security. The architecture combining Redis, Kafka, RabbitMQ, MongoDB, PostgreSQL, OAuth2 and JWT is well designed and can withstand and process transactions efficiently at a high rate. The system's ability to achieve low latencies while maintaining high reliability and data consistency is a great enabler in improving the travel booking system, at both the customer and business level.

## 7. Discussion

The microservices oriented architecture presented in this airline reservation system is very effective when designing travel applications which have high requirements. The system uses caching leveraging Redis, real time messaging with Kafka and RabbitMQ, data storage using MongoDB and PostgreSQL alongside OAuth2 with JWT for secure and efficient authentication, thereby achieving performance, scalability and security at its best. The very high data integration rate (99.5%) and the low inter service latency (less than 100 ms) speaks to the architecture's effectiveness in synchronizing service's data. In addition to this, the responsive nature of the system was aided by a very high cache hit rate (92%) thereby reducing the load on the database which was beneficial in peak periods. Because the high volume transactions occurred, the system was designed to be scalable via the use of Docker and Kubernetes thus allowing elasticity. The systems low error rates (0.2%) was also an indicator of the stability and fault tolerance of the architecture. To sum up, this architecture corresponds to the standards applied in designing a stout airline reservation system, ensuring a broad spectrum of functionality while providing safety to the users and their resources as well as room for further development.

## 8. Conclusion

To summarize, the complex system of microservices which has been designed for an airline reservation system fulfills the crucial requirements of elasticity, data integration and safety which are normally found in any advancement travel application. It included, Redis for caching, Kafka, RabbitMQ for real-time messaging, MongoDB and PostgreSQL for data storage optimization all secured by JWT authentication over OAuth2 therefore the system was both high performing and highly reliable. Some key statistics that highlight the system architecture such as data consistency which stood at 99.5%, cache hit rate at 92 % and average data propagation latency of 100 milliseconds or lower clearly shows the system does well in keeping the data accurately while less stressing the database and responding in time. The system designed for scalability using a 3-tier architecture containerized with Docker and orchestrated with Kubernetes allows easy horizontal scaling and maximum throughput of 1050 calls per second during peak seasons with the error rate just 0.2%. This implies that the design meets the needs presented by airline reservations systems in terms of security, performance and scalability, hence ease of use and business growth is guaranteed. With particular regard to the trends within the industry, further developments could also aim at integrating deeper AI-based user experience and recommendations systems.